\documentclass[10pt,twocolumn,letterpaper]{article}

\usepackage[pagenumbers]{cvpr} 
\usepackage{csquotes}
\usepackage[ruled,vlined]{algorithm2e}

\usepackage[abs]{overpic}
\usepackage[dvipsnames, table]{xcolor}

\definecolor{cvprblue}{rgb}{0.21,0.49,0.74}
\usepackage[pagebackref,breaklinks,colorlinks,citecolor=cvprblue]{hyperref}

\newcommand{\thinparagraph}[1]{\vspace{0.1em}\par\noindent\textbf{#1}\enspace}

\newcommand\blfootnote[1]{%
  \begingroup
  \renewcommand\thefootnote{}\footnote{#1}%
  \addtocounter{footnote}{-1}%
  \endgroup
}

\def\paperID{****} 
\def\confName{CVPR}
\def\confYear{2024}

\title{Image Sculpting: Precise Object Editing with 3D Geometry Control}

\author{Jiraphon Yenphraphai\textsuperscript{1} \quad
Xichen Pan\textsuperscript{1} \quad
Sainan Liu\textsuperscript{2} 
\quad
Daniele Panozzo\textsuperscript{1} 
\quad
Saining Xie\textsuperscript{1} \\[3mm]
\textsuperscript{1}New York University \qquad \textsuperscript{2}Intel Labs}

\begin{document}
\twocolumn[{
\maketitle
\vspace{-37pt}
\begin{center}
    \centering
    \captionsetup{type=figure}
    \includegraphics[width=\linewidth]{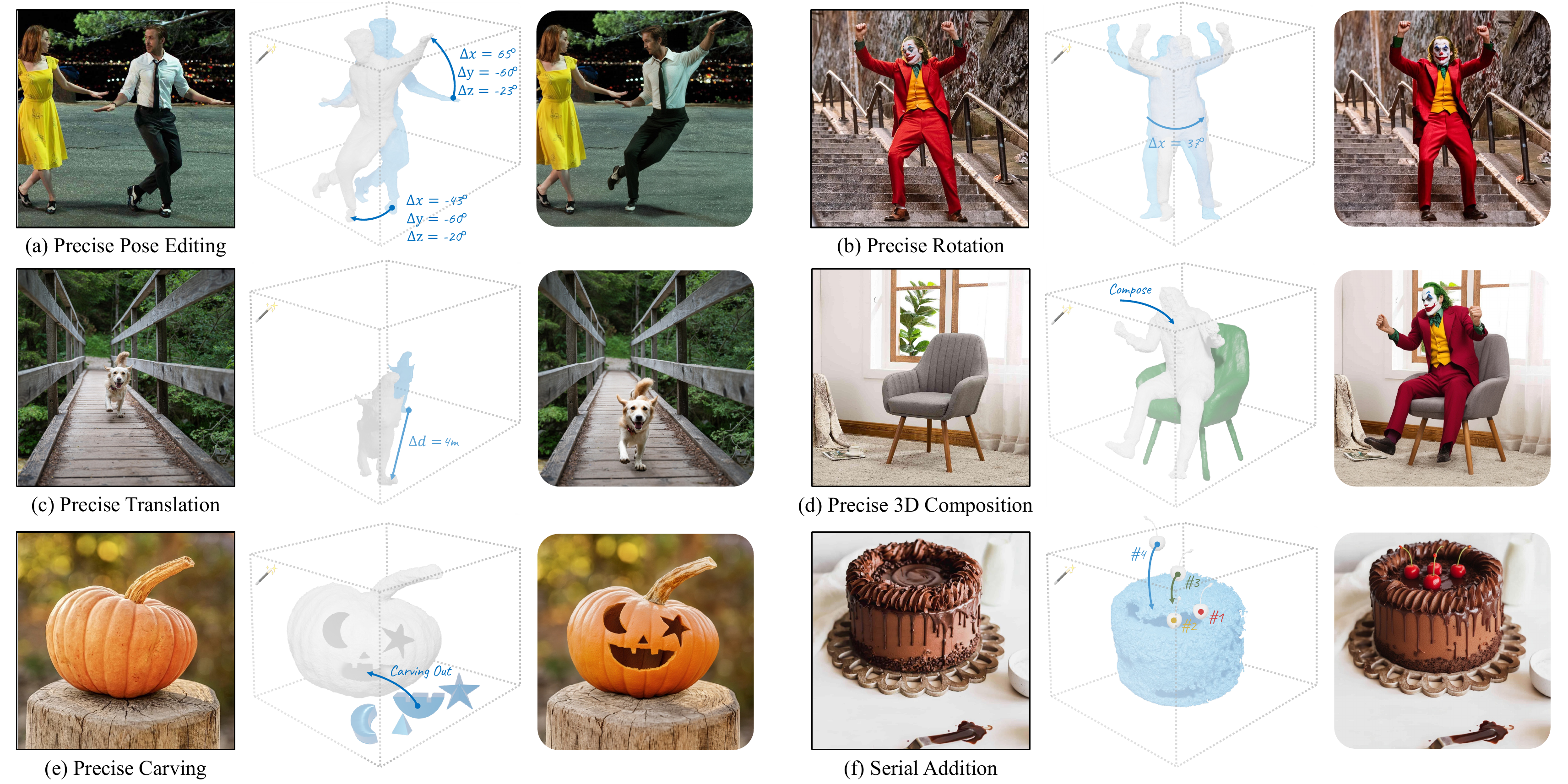}
    \captionof{figure}{Achieving \emph{precise} control in image editing tasks can be challenging with standard 2D generative pipelines. Our \emph{Image Sculpting} framework offers the ability to interact with 3D geometry starting with a single image. This enables users to perform detailed, quantifiable, and physically-plausible edits, including \emph{precise} pose editing, rotation, translation, 3D composition, carving, and serial addition.}
    \label{fig:teaser}
\end{center}
}]
\maketitle
\begin{abstract}
\vspace{-0.5em}
We present \emph{Image Sculpting}, a new framework for editing 2D images by incorporating tools from 3D geometry and graphics. This approach differs markedly from existing methods, which are confined to 2D spaces and typically rely on textual instructions, leading to ambiguity and limited control. Image Sculpting converts 2D objects into 3D, enabling direct interaction with their 3D geometry. Post-editing, these objects are re-rendered into 2D, merging into the original image to produce high-fidelity results through a coarse-to-fine enhancement process. The framework supports precise, quantifiable, and physically-plausible editing options such as pose editing, rotation, translation, 3D composition, carving, and serial addition. It marks an initial step towards combining the creative freedom of generative models with the precision of graphics pipelines.\blfootnote{\ \ \ Code and project page available \href{https://image-sculpting.github.io/}{here}.}
\end{abstract}

\section{Introduction}
Recent developments in the field of image generative modeling~\cite{dalle2, imagen, parti, ldm} have unlocked new potentials in creative content creation, offering unprecedented opportunities for the generation of diverse visual content by materializing ideas and concepts articulated through language prompts. However, the integration of these models into real-world content creation workflows still poses significant challenges. Among the most critical is the need for users to have detailed control over various aspects of generated objects, including their pose, shape, location, layout, and spatial compositions. The precision extends to quantifiable manipulations, such as rotating an object by a specific angle or making physically-realistic modifications, such as positioning a character in a way that conforms to basic anatomical and physical principles. Interestingly, such a quest for precision and controllability aligns closely with the core principles of computer graphics, which strive to generate photorealistic images with artistic control.

\begin{figure*}[!t]
    \centering
    \begin{subfigure}[t]{0.42\linewidth}
        \includegraphics[height=8em]{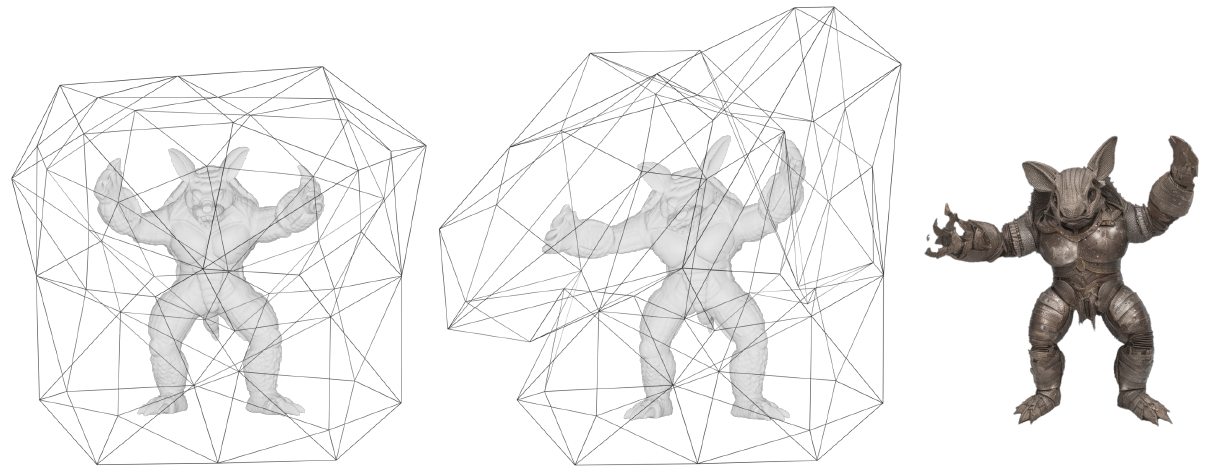}
        \caption{Space Deformations}
    \end{subfigure}
    \begin{subfigure}[t]{0.285\linewidth}
        \includegraphics[height=8em]{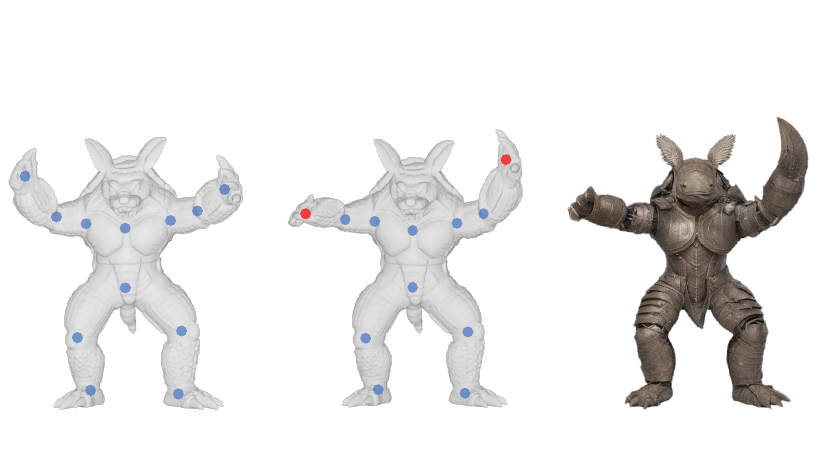}
        \caption{Shape-Aware Deformation}
    \end{subfigure}
    \begin{subfigure}[t]{0.285\linewidth}
        \includegraphics[height=8em]{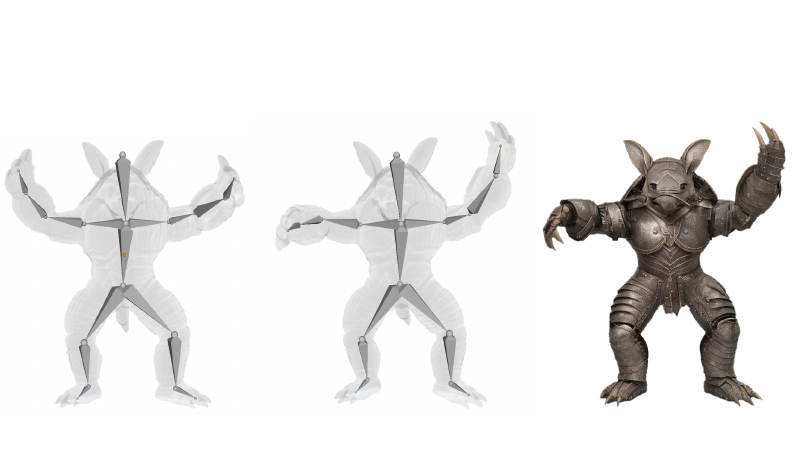}
        \caption{Linear Blend Skinning}
    \end{subfigure}
    \caption{Illustration of three mesh deformation methods applied to a 3D model. In cage-based space deformation (a), the model is placed in a cage and deformed when the user moves the cage vertices \cite{Ju:2005}. As-Rigid-As-Possible (ARAP)~\cite{ARAP_modeling:2007} deformation (b) deforms the model when user-selected blue handle points are moved towards designated red target points. Linear blend skinning (c) maps the deformation of a skeleton to the model ~\cite{skinningcourse:2014}. Following deformation, a diffusion rendering process can be added for controllable generation. Each mesh deformation technique offers a different balance of control, speed, and precision. Our framework can use any of these techniques.}    
    \label{fig:cbd_arap_lbs}
\end{figure*}

In virtual effects (VFX) and rendering pipelines, experts meticulously craft and edit every detail within a fully controllable \emph{3D environment}, striving for utmost realism.
For decades, methods for accurately manipulating and rendering objects have been explored, leading to the development of numerous advanced techniques in 3D model acquisition, rigging, posing, lighting, texturing, and scene rendering. These methods form the bedrock of the modern computer graphics pipeline. However, it often requires custom hardware and software for (1) acquiring production-quality 3D models or designing them from scratch, (2) making these models possible to animate (rigging), (3) creating visually plausible animations (animation), (4) rendering back in the 2D world after applying material and setting up the lighting, and (5) compositing the resulting image with a background or other objects. This process often employs teams of artists and engineers for each one of these steps, as it requires substantial manual input using specialized tools (\eg After Effects~\cite{aftereffects2023}, Substance~\cite{substance2023}, and 3ds Max~\cite{autodesk3dsmax2023}). 

In contrast, AI-based image generation avoids all this manual work, requiring only a text prompt. Leveraging the power of human language and large datasets of curated content, transforming a text description into a visually striking image is more accessible than ever. Yet, when it comes to precise object manipulation, the current 2D-based generative approach faces inherent limitations due to the lack of a third dimension, leading to incomplete information, limited user interaction on a flat plane, and possible ambiguities. The gap in controllability with respect to image generation using computer graphics techniques is striking, and closing it is a major goal of our work.

Most interfaces for image editing frameworks rely on text-based instructions. For example, techniques such as Prompt-to-Prompt~\cite{prompt2prompt}, Plug-and-Play~\cite{pnp}, InstructPix2Pix~\cite{instructpix2pix}, Imagic~\cite{imagic} and Object 3DIT~\cite{3dit} offer adaptable language control. However, achieving precise manipulation through these models remains a challenge. Straight-forward manipulations such as \emph{``changing a style to mimic Van Gogh''} are manageable. However, more specific instructions such as \emph{``lift the object by 5 cm and rotate it by 42 degrees.''} are less likely to be successful, as current generative models cannot fulfill such detailed requests through textual prompts alone. 2D-based interactive methods such as DragGAN~\cite{draggan}, FreeDrag~\cite{freedrag}, and DragDiffusion~\cite{dragdiffusion} demonstrate the ability to alter part of an object through transitions in the latent space. Despite this, they have their limitations: 1) they can accomplish basic deformations, but the outcomes are not entirely predictable, often leading to results that do not align with the user's intentions; 2) these latent transformations operate within the 2D feature space, which inherently limits their ability to represent 3D transformations and handle occlusions accurately; 3) they lack physics-awareness, which complicates incorporating external constraints, such as skeletal structures.

Our work draws inspiration from the computer graphics pipeline and ventures into a novel approach for 2D image-based object manipulation tasks. Our proposed \emph{Image Sculpting} framework, which metaphorically suggests the flexible and precise sculpting of a 2D image in a 3D space, integrates three key components: (1) single-view 3D reconstruction, (2) manipulation of objects in 3D, and (3) a coarse-to-fine generative enhancement process. More specifically, 2D objects are converted into 3D models, granting users the ability to interact with and manipulate the 3D geometry directly, which allows for precision in editing. The manipulated objects are then seamlessly reincorporated into their original 2D contexts, maintaining visual coherence and fidelity. A critical hurdle in this process is the single-view 3D reconstruction method, a task that, despite rapid progress~\cite{zero123, qian2023magic123, liu2023one, shi2023zero123++, hong2023lrm, lin2023consistent123, liu2023syncdreamer, long2023wonder3d}, often results in relatively low-fidelity, coarse geometric and texture representations. Unlike manually crafted 3D assets used for graphics, their rendered version is far from photo-realistic. Nonetheless, the extracted geometries are sufficient for interactive and precise control. To achieve high-quality final images, a separate enhancement procedure is necessary. In summary, our Image Sculpting pipeline has three key phases:

\textbf{Phase 1.} For the 3D reconstruction phase, we employ a zero-shot single image reconstruction model (Zero-1-to-3~\cite{zero123}), which has been trained on extensive datasets~\cite{deitke2023objaverse} of 3D objects.

\textbf{Phase 2.} The deformation process utilizes established geometric processing tools, such as As-Rigid-As-Possible (ARAP)~\cite{ARAP_modeling:2007} and linear-based skinning~\cite{Thalmann:1989}, enabling interactive and precise manipulation of the 3D models. 

\textbf{Phase 3.} For the generative enhancement process, we develop a coarse-to-fine enhancement approach, using an improved feature injection technique~\cite{pnp}. Our method strikes a balance between maintaining the original texture of the object and the modified geometry, utilizing a pre-trained text-to-image diffusion model with additional controls.

Our Image Sculpting framework showcases an array of precise and quantifiable image editing capabilities. These include precise pose editing, rotation, translation, multi-object 3D composition, carving, and serial addition. This suite of functionalities demonstrates the versatility of our approach and its superiority in precision and control compared to existing image editing methods. Our method also outperforms various baselines in image quality, as confirmed by both qualitative and quantitative evaluations on the new \emph{SculptingBench} benchmark. We believe that our method can foster new opportunities in merging the flexibility of generative models with the precise controllability inherent in traditional graphics pipelines.

\section{Related Work}
\thinparagraph{Generative Image Editing}
In computer graphics, extensive research on interactive raster image editing exists, and we defer its detailed review to the next section. In computer vision, the advent of image generative models such as GANs~\cite{gan, karras2019style, karras2020analyzing} has expanded the scope of image editing to include style transfer~\cite{styletransfer}, image-to-image translation~\cite{pix2pix,zhu2017unpaired}, latent manipulation~\cite{wu2021stylespace, shen2020interpreting}, and text-based manipulation~\cite{styleclip, xia2021tedigan,abdal2022clip2stylegan}. Recently, capabilities in image editing have advanced significantly with the rise of diffusion models~\cite{adm, ldm, dit}. The leading systems~\cite{dalle2, dalle3, imagen, midjourney} allow users to generate image variations or use inpainting masks~\cite{nichol2022glide} to generate specific parts of scenes based on a text prompt. Other work explores enhancing pre-trained diffusion models with text-guided editing capabilities~\cite{instructpix2pix, pnp, prompt2prompt, nulltextinversion}. Yet, text-based editing has limitations in precisely controlling object shapes and positions. ControlNet~\cite{controlnet} incorporates additional conditional inputs such as depth~\cite{ranftl2020towards}, poses~\cite{openpose}, and edges~\cite{hed} for controllable generation. For more intuitive interactions, DragGAN~\cite{draggan} enables users to drag control points on objects with GANs, and similar techniques have been adapted for diffusion models~\cite{freedrag, dragdiffusion}. However, these methods are mostly confined to 2D and face challenges in tasks requiring more complex, out-of-plane transformations. 3D-aware generative models such as EG3D~\cite{Chan2021} and StyleNeRF~\cite{gu2021stylenerf} have explored this direction. OBJect-3DIT~\cite{3dit}, a baseline in our paper, studied 3D-aware editing using language instructions. However, its effectiveness is somewhat constrained due to its training on a synthetic dataset.

\thinparagraph{Single-View Reconstruction}
Single-view 3D reconstruction is a long-standing problem in computer vision~\cite{hartley2003multiple}. While algorithmic advancements are important, the significance of training data has been increasingly recognized. Earlier efforts were geared towards training models~\cite{sitzmann2019scene, niemeyer2020differentiable, yu2021pixelnerf,wu2023multiview} using smaller, simplistic 3D datasets~\cite{chang2015shapenet, reizenstein2021common}. Recent approaches~\cite{dreamfusion, scorejac} 
have started to utilize density distillation from pre-trained 2D diffusion models trained on large-scale text-image datasets, lessening the reliance on 3D data. Nonetheless, for improved view-consistency, the demand for high-quality 3D data is indispensable. The emergence of large-scale 3D datasets, such as Objaverse~\cite{deitke2023objaverse, deitke2023objaversexl}, has spurred methods such as Zero-1-to-3~\cite{zero123} to combine 2D score distillation with 3D data training. This has led to a surge in new models in this domain, noticeably enhancing reconstruction quality~\cite{qian2023magic123,liu2023one,wang2023prolificdreamer,shi2023MVDream}.
Current 3D reconstruction models, while not perfect, have attained a level of maturity that makes them suitable for shape editing.

\section{Overview of 3D Shape Deformation}
\label{sec:prelim}
\begin{figure*}[!t]
    \centering
    \includegraphics[width=\linewidth]{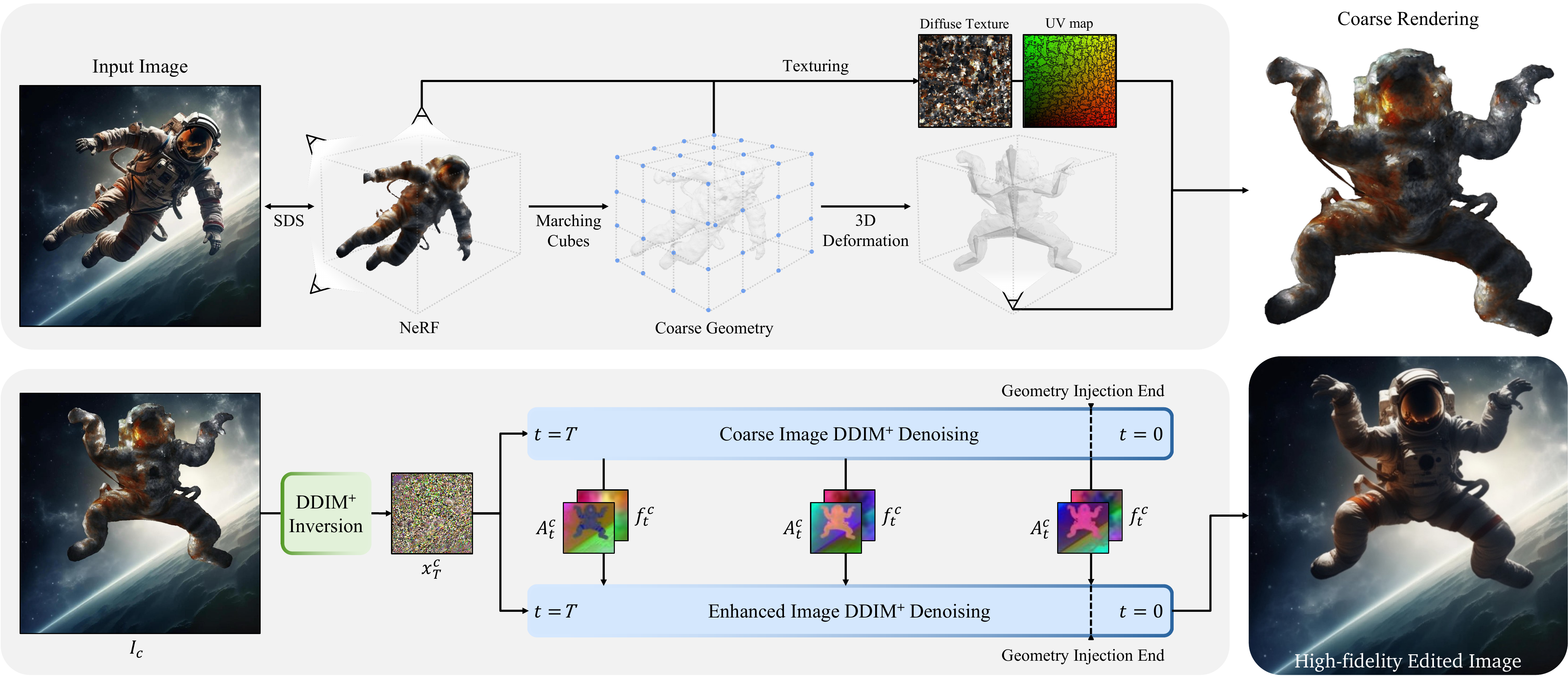}
    \caption{Overview of our \emph{Image Sculpting} pipeline, DDIM$^\textbf{+}$ represents DDIM with the DreamBooth fine-tuned and depth controlled model. The process begins by converting the input image into a textured 3D model through a de-rendering process.
    This model is then prepared for interactive deformation by creating a skeleton and calculating skinning weights. The user can modify the skeleton to deform the model, resulting in an initial coarse image. To refine this edited image, we invert the coarse rendering $I_c$ into the noise $\boldsymbol{x}_T^c$. We then inject self-attention maps $\boldsymbol{A}_t^c$ and feature maps $\boldsymbol{f}_t^c$ 
    from the initial image's denoising process into the enhanced image denoising steps. This technique helps in preserving the geometry of the modified object while restoring the visual quality of the edited image.}
    
    \label{fig:algo}
    \vspace{-1em}
\end{figure*}

The deformation of 3D shapes has been extensively studied in the last four decades, with both traditional and data-driven methods being proposed and successfully used in robotics, graphics, and engineering. We review the main approaches and their usability within our framework.

\thinparagraph{Space Deformations}
The older and still widely used approach is applying a volumetric warp function $f: R^3 \rightarrow R^3$ to all points of a 3D domain \cite{Sederberg:1986}. This approach can be applied to explicit (triangular or polygonal meshes) or implicit representations. The map can be parametrized using splines on lattices \cite{Sederberg:1986}, vertices on a cage \cite{Ju:2005}, or neural fields \cite{Dodik:VBC:2023}. A limitation of these approaches is that they are unaware of the object shape, making them more challenging to use on complex articulated objects \cite{PMP:2010}.

\thinparagraph{Shape-Aware Deformation}
Shape-aware deformations provide a set of controls linked to the objects' surface. In Computer-Aided-Design (CAD), a small set of control points define a smooth surface using spline patches \cite{Farin:2001}. Despite its flexibility and quality, extracting spline patches from  3D models or NeRFs is a challenging and open problem \cite{Bommes:2012:SAQ}. Partial differential equation (PDE)-based methods simulate the deformation of an object, representing it as a volumetric deformable solid \cite{Sifakis:2012} or as a thin rubber shell \cite{ARAP_modeling:2007}. The forces guiding the deformation are applied by moving handles selected on a surface \cite{Botsch:2008}, making them intuitive to use and requiring minimal user interaction.

\thinparagraph{Linear Blend Skinning}
The most popular deformation approach is linear blend skinning \cite{skinningcourse:2014}, which defines a space deformation function as a blended average of a set of affine transformations weighted by shape-aware scalar functions, often computed with methods based on solutions of PDEs on surfaces \cite{BBW:2011} or manually edited. This approach offers complete control and flexibility, as the affine transformation can be attached to points, vertexes of a cage, or segments in a skeleton \cite{baran2007automatic}. 

\thinparagraph{Our approach} We can use any of these algorithms to precisely control the shape deformation and, thus, the rendered image. We show an example of one representative method for each class in Fig~\ref{fig:cbd_arap_lbs}, and we leave as future work additional automation of this step.

\section{Methods}
Given a single 2D image, our objective is to enable precise manipulation of the objects and their orientations in 3D space, before converting this back into a high-quality edited 2D image. To achieve this, we have developed a novel editing pipeline tailored for image sculpting (see Fig~\ref{fig:algo}) composed of three steps: (1) We initially convert the input image into a 3D model, (2) the 3D model is edited by deforming it in 3D space, and (3) we use a coarse-to-fine generative enhancement pipeline to turn the coarse rendering of the 3D model into a high-fidelity image.

\subsection{De-Rendering and Deformation}
\label{sec:methods_3d}
Given an image of an object, our goal is to perform 3D reconstruction to obtain its 3D model. 

\thinparagraph{Image to NeRF} With advancements in text-to-image foundation models~\cite{ldm} and the viewpoint-conditioned image translation model~\cite{zero123}, our initial step involves segmenting the selected object from the input image using SAM ~\cite{kirillov2023segment}. Building upon this, we then train a NeRF using Score Distillation Sampling (SDS)~\cite{dreamfusion}.

\thinparagraph{NeRF to 3D Model} 
We use the implementation in threestudio~\cite{threestudio2023} to convert a NeRF volume into a mesh. This algorithm transforms the volume density into a signed distance function, extracts an isosurface \cite{marchingcubes}, and parameterizes it \cite{xatlas} for texture mapping \cite{Shirley:2009}. The texture is extracted by differentiable rendering \cite{Laine2020diffrast}.

\begin{figure*}[t]
    \centering
    \begin{overpic}[width=\linewidth]{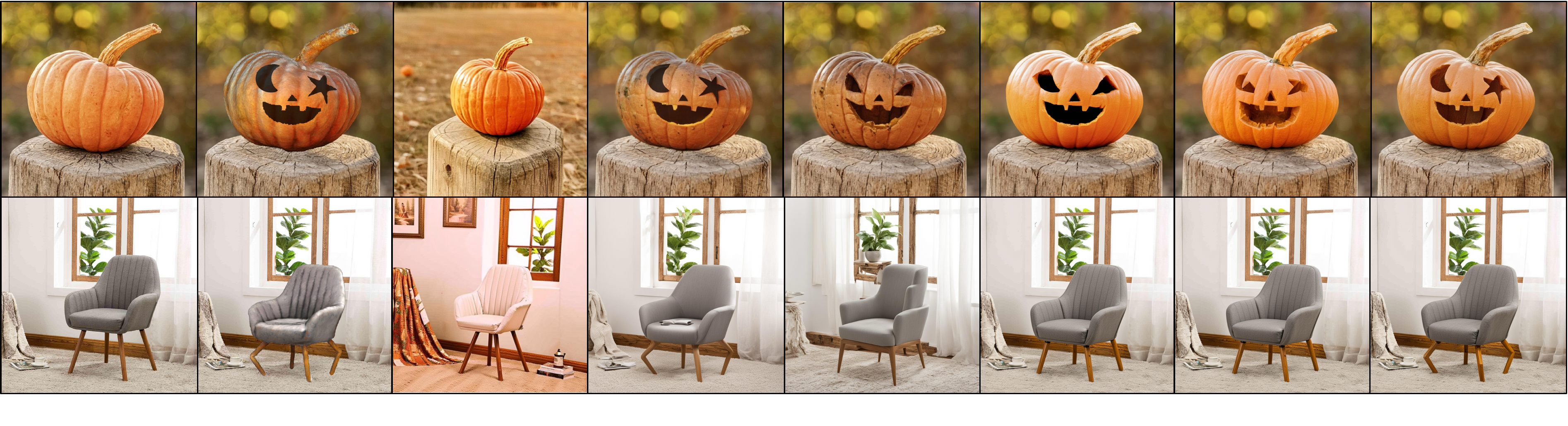}
    \put(16,4) {\tiny Input Image}
    \put(76,4) {\tiny Coarse Rendering}
    \put(137,4) {\tiny DreamBooth~\cite{dreambooth}}
    \put(191,4) {\tiny SDEdit~\cite{meng2021sdedit} ($t_0=0.4$)}
    \put(253,4) {\tiny SDEdit~\cite{meng2021sdedit} ($t_0=0.6$)}
    \put(315,4) {\tiny Ours w/o Feature Injection}
    \put(380,4) {\tiny Ours w/o Depth Control}
    \put(460,4) {\tiny Ours}
    \end{overpic}
    \vspace{-1.65em}
    \caption{Comparison of our final method with various baseline methods and ablations. Our approach effectively maintains the geometric information while ensuring the texture quality. In contrast, other methods typically preserve either the texture or the geometry, but not both.}
    \label{fig:c2f_comparison}
\end{figure*}
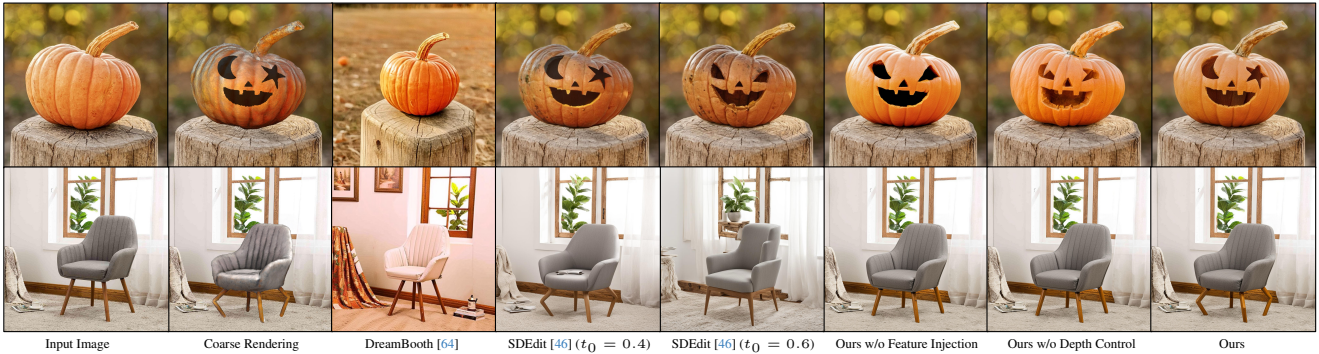
\thinparagraph{3D Model Deformation}
After obtaining the 3D model, a user can manually construct a skeleton and interactively manipulate it by rotating the bones to achieve the target pose. The mesh deformation affects the vertex positions of the object but not the UV coordinates used for texture mapping; this procedure thus deforms the texture mapped on the object following its deformation.

However, the resulting image quality depends on the 3D reconstruction's accuracy, which, in our case, is coarse and insufficient for the intended visual outcome (Fig~\ref{fig:algo}). Therefore, we rely on an image enhancement pipeline to convert the coarse rendering into a high-quality output.

\begin{figure}[!t]
    \centering
    \includegraphics[width=\linewidth]{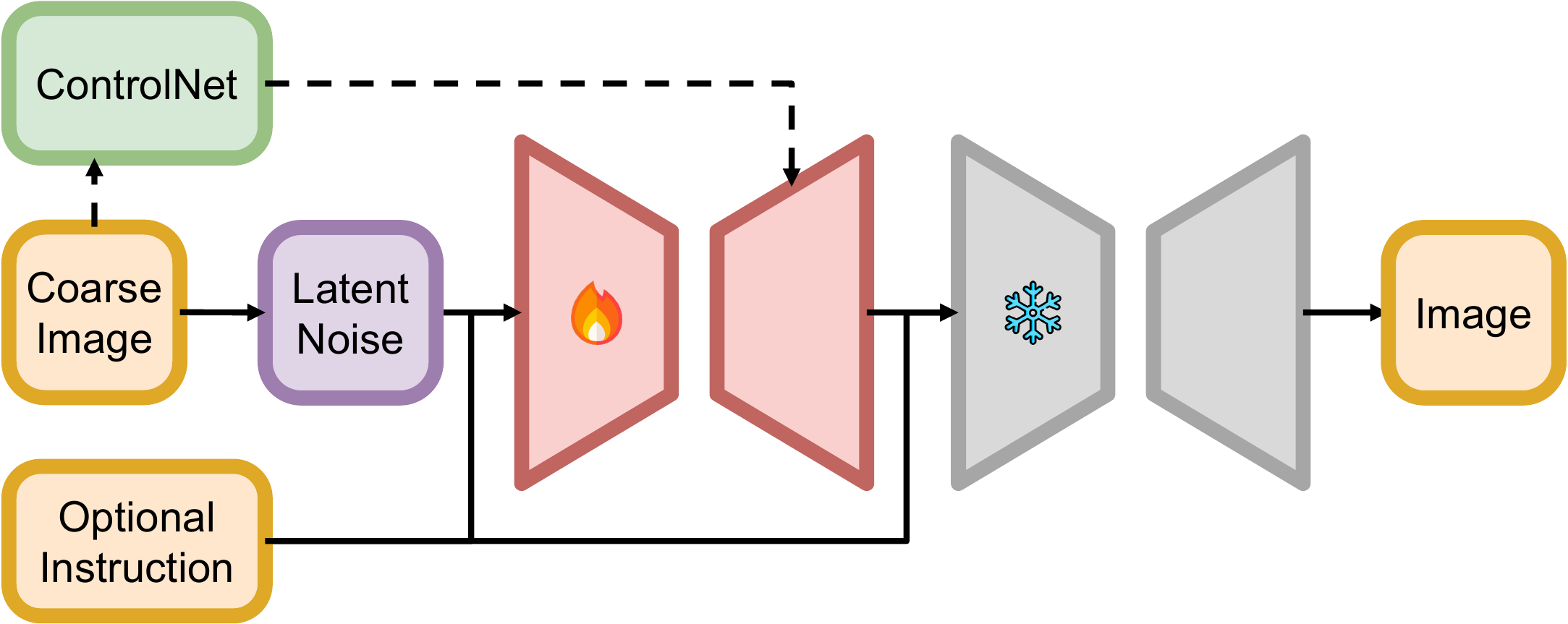}
    \caption{Overview of the coarse-to-fine generative enhancement model architecture. The red module denotes the one-shot DreamBooth~\cite{dreambooth}, which requires tuning; the grey module is the SDXL Refiner~\cite{refiner}, which is frozen in our experiments.}
    \label{fig:arch}
    \vspace{-1em}
\end{figure}

\begin{figure*}[t]
    \centering
    \includegraphics[width=0.95\linewidth]{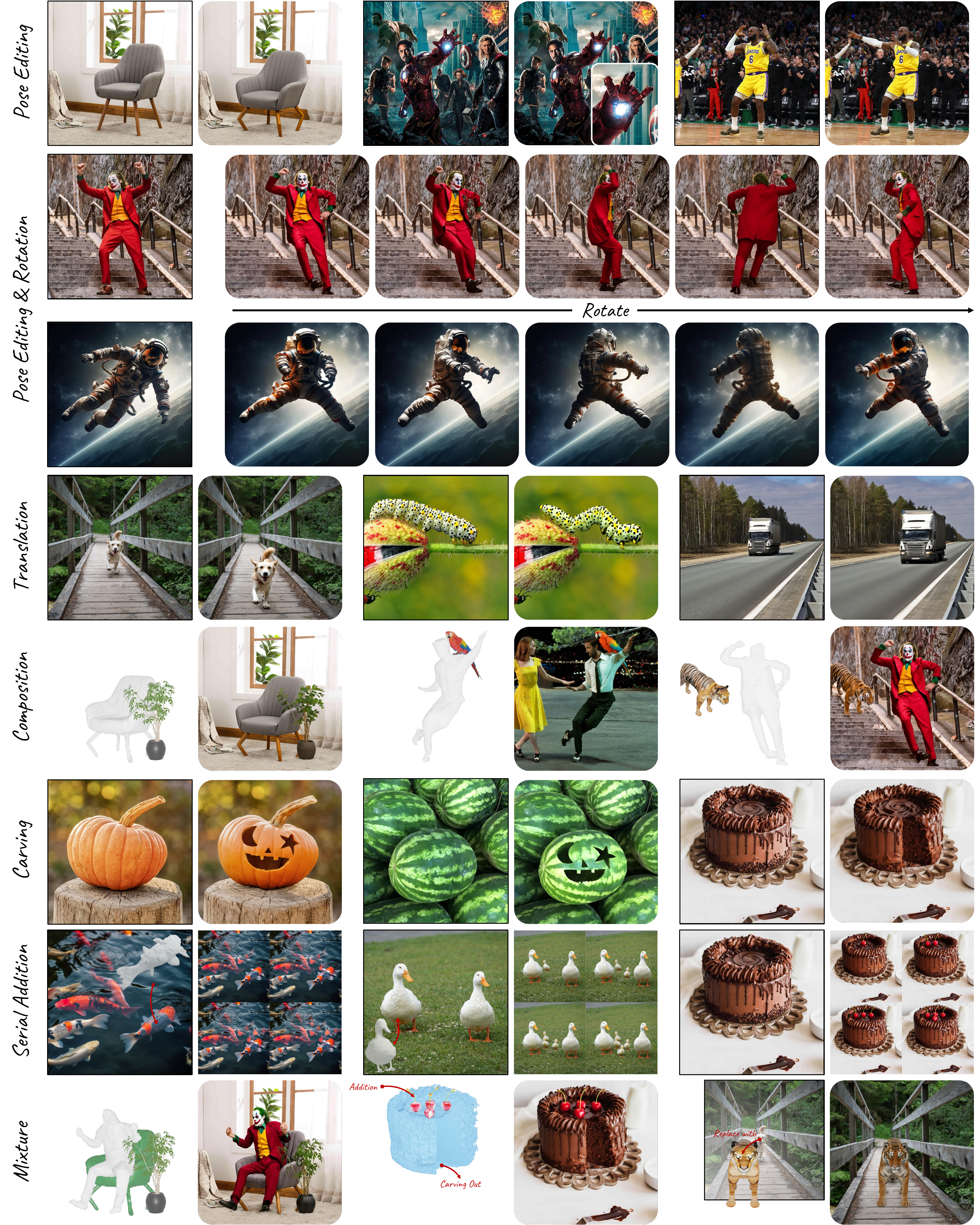}
    \caption{A compilation of qualitative results from six image editing tasks. Additionally, we include additional examples (termed as `\textit{Mixture}' in the final row) to illustrate the versatile combination of these capabilities.}
    \label{fig:results}
\end{figure*}

\subsection{Coarse-to-Fine Generative Enhancement}
\label{sec:coars-to-fine}
This section focuses on blending a coarsely rendered image back to its original background. The aim is to restore textural details while keeping the edited geometry intact. 
Image restoration and enhancement are commonly approached as image-to-image translation tasks~\cite{wang2020deep}, leveraging the strong correlation between the source and target images. 
Our challenge, however, presents a unique scenario: despite overall similarities in appearance and texture between the input and desired output, the input object's geometry changes, sometimes significantly, after user editing. 

In exploring possible solutions, one approach is to use subject-driven personalization techniques like DreamBooth~\cite{dreambooth}. They aim to preserve key details from the input, but might compromise the edited geometry. Alternatively, image-to-image translation methods like SDEdit~\cite{meng2021sdedit} can be used to preserve the edited geometry, but this might disturb the textural consistency with the original image. This dichotomy was clear in our preliminary study, as shown in Fig~\ref{fig:c2f_comparison}. SDEdit can maintain the geometry, but it was unable to accurately replicate the textures. On the other hand, DreamBooth produced high-fidelity outputs, but struggled to preserve both the texture and geometry effectively. 

To address the balance between preserving texture and geometry, our approach begins by ``personalizing'' a pre-trained text-to-image diffusion model. To capture the object's key features, we fine-tune the diffusion model with DreamBooth on \emph{one} input reference image. To maintain the geometry, we adapt a feature and attention injection technique~\cite{pnp}, originally designed for semantic layout control. Furthermore, we incorporate depth data from the 3D model through ControlNet~\cite{controlnet}. We find this integration crucial in minimizing uncertainties during the enhancement process.

\thinparagraph{One-shot Dreambooth}
DreamBooth~\cite{dreambooth} fine-tunes a pre-trained diffusion model with a few images for subject-driven generation. 
The original DreamBooth paper~\cite{dreambooth} has shown its ability to leverage the semantic class priors to generate novel views of an object, given only a few frontal images of the subject. This aspect is particularly useful in our setting, since the coarse rendering we work with lacks explicit viewpoint information. In our application, we train DreamBooth using just a single example, which is the input image. Notably, this one-shot approach with DreamBooth also effectively captures the detailed texture, thereby filling in the textural gaps present in the coarse rendering.

\thinparagraph{Depth Control}
We use depth ControlNet~\cite{controlnet} to preserve the geometric information of user editing. The depth map is rendered directly from the deformed 3D model, bypassing the need for any monocular depth estimation. For the background region, we don't use the depth map. This depth map serves as a spatial control signal, guiding the geometry generation in the final edited images. However, relying solely on depth control is not sufficient -- although it can preserve the geometry to some extent, it still struggles in local, more nuanced editing, such as capturing the specific shapes of a pumpkin's eyes or the bent legs of a chair (Fig~\ref{fig:c2f_comparison}).

\thinparagraph{Feature Injection}
To better preserve the geometry, we use feature injection. As demonstrated in Fig~\ref{fig:algo}, this step begins with DDIM inversion~\cite{song2020denoising} (with the DreamBooth finetuned, depth controlled diffusion model) of the coarse rendering image to obtain the inverted latents. At each denoising step, we denoise the inverted latent of the coarse rendering along with the latent of the refined image, extracting their respective feature maps (from the residual blocks) and self-attention maps (from the transformer blocks). It has been shown in~\cite{pnp} that the feature maps carry semantic information, while the self-attention maps contain the geometry and layout of the generated images. By overriding the feature and self-attention maps during the enhanced image denoising steps with those from the coarser version, we ensure the geometry of the enhanced image can reflect those of the coarse rendering. The pseudo code for our generative enhancement is detailed in Appendix \ref{alg:1}. Note that our method differs from the original Plug-and-Play use cases: we use feature injection to preserve the geometry during the coarse-to-fine process rather than translating the image according to a new text prompt. We present the injection layer selection and the replacement schedule in Section~\ref{sec:exp}.

\begin{figure}[!t]
    \centering
    \includegraphics[width=\linewidth]{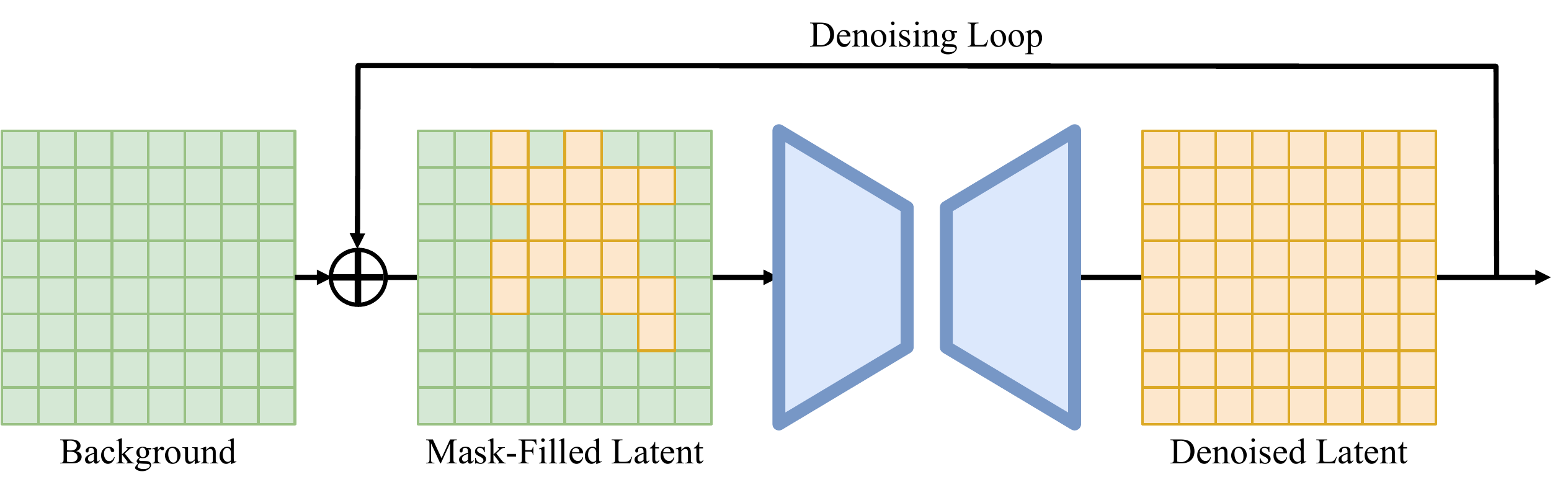}
    \vspace{-1.65em}
    \caption{Our blend-in process. At every denoising step, we mask the background areas and blend them with the unmasked regions from the denoised latent. This process helps maintain visual coherence and preserve the background.}
    \label{fig:blendin}
    \vspace{-1em}
\end{figure}
\label{sec:bg_harmonization}

\begin{figure}[!t]
    \centering
    \includegraphics[width=\linewidth]{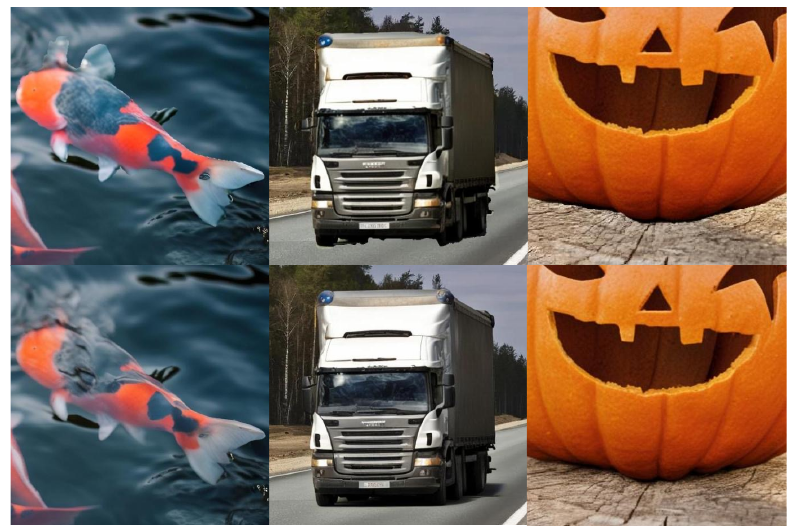}
    \vspace{-1.9em}
    \caption{Our blend-in process yields visually harmonious results. \emph{Top:} Results from direct copy-pasting. \emph{Bottom:} Our results.}
    \label{fig:bg_blend_in}
    \vspace{-1.5em}
\end{figure}

\thinparagraph{Background Blend-In} To maintain the consistency of the background between the input and edited images, we first inpaint the area initially occupied by the object in the input image, thus obtaining an unobstructed background. However, another challenge arises in merging the edited object into this background smoothly. Merely copy-pasting it onto the background leads to an unrealistic visual effect, such as the improper water reflections over the fish and the absence of shadow casting from the truck (Fig~\ref{fig:bg_blend_in}). To overcome this, as demonstrated in Fig~\ref{fig:blendin}, our approach involves masking the background areas during the denoising steps to preserve their original background. This means we retain the unedited background by blending the unmasked (edited) regions from the denoising step with the masked (original) background. We use SDXL~\cite{podell2023sdxl} as our pre-trained text-to-image model, which includes a refiner module by default. We keep this module in our pipeline, as empirically it slightly enhances the results by reducing artifacts.

\section{Experiments}
\label{sec:exp}
\thinparagraph{Experimental Setup}
we follow~\cite{threestudio2023} to obtain the initial NeRF representation and to extract the textured 3D model. We use Instant-NGP~\cite{muller2022instant} and a grid size of 256 for the 3D model extraction from NeRF. During the coarse-to-fine generative enhancement process, for one-shot DreamBooth, we fine-tune the SDXL-1.0~\cite{podell2023sdxl} model using LoRA \cite{hu2021lora} for 800 steps with a learning rate of 1e-5. For feature injection stage, we utilize all the self-attention layers of the SDXL decoder and the first block of the SDXL's upsampling decoder. We set $\tau_A=0.5$ and $\tau_f=0.2$. The SDXL refiner is applied after $t=0.1T$. For background inpainting, we use Adobe generative fill~\cite{firefly}. 

\begin{figure}[!t]
    \centering
    \begin{overpic}[width=\linewidth]{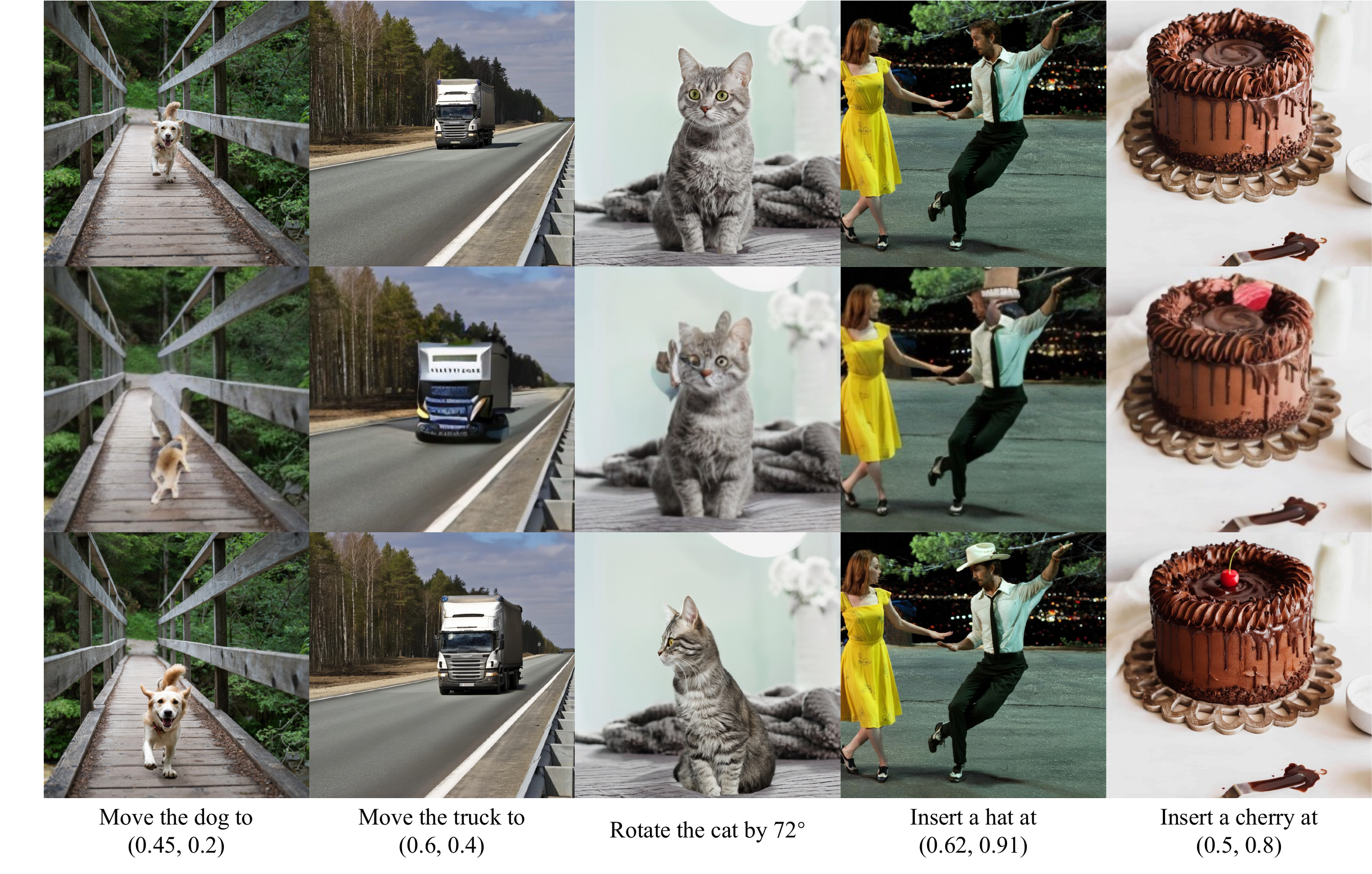}
    \put(0,30) {\rotatebox{90}{\tiny Ours}}
    \put(0,72) {\rotatebox{90}{\tiny 3DIT~\cite{3dit}}}
    \put(0,116) {\rotatebox{90}{\tiny Input Image}}
    \end{overpic}
    \vspace{-1.95em}
    \caption{Comparisons with OBJect-3DIT~\cite{3dit} on object translation, rotation, and composition tasks.}
    \label{fig:baseline_comparison_obj3dit}
    \vspace{-0.5em}
\end{figure}

\begin{figure}[!t]
    \begin{overpic}[width=\linewidth]{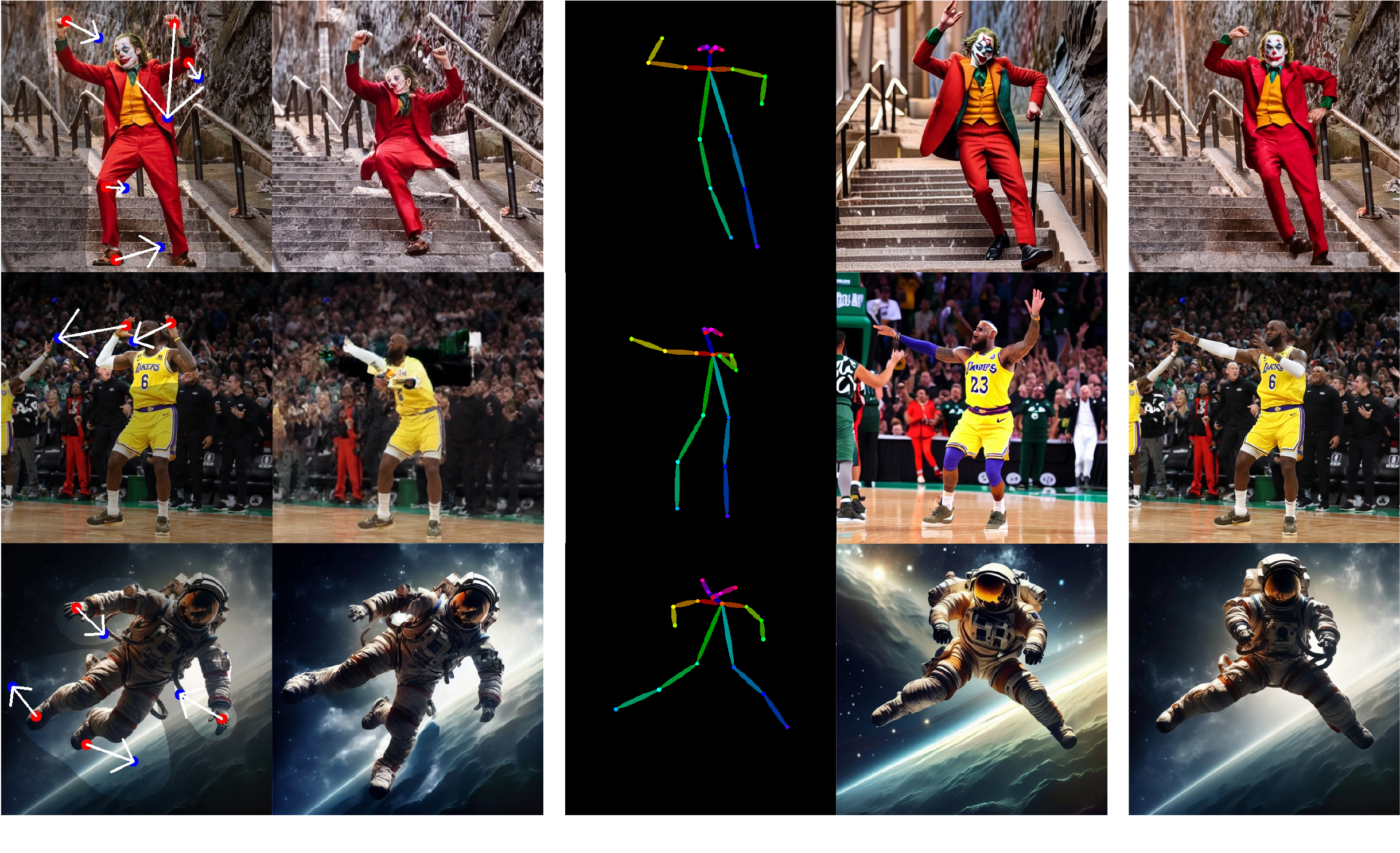}
    \put(11,0) {\tiny Drag Control}
    \put(51,0) {\tiny DragDiffusion~\cite{dragdiffusion}}
    \put(101,0) {\tiny Openpose Control}
    \put(150,0) {\tiny ControlNet~\cite{controlnet}}
    \put(210,0) {\tiny Ours}
    \end{overpic}
    \vspace{-1.65em}
    \caption{Comparisons with DragDiffusion~\cite{dragdiffusion} and ControlNet~\cite{controlnet} on pose editing. These techniques face difficulties in handling complex pose modifications.}
    \label{fig:baseline_comparison_pose}
    \vspace{-1.2em}
\end{figure}

\begin{figure}[!t]
    \begin{overpic}[width=\linewidth]{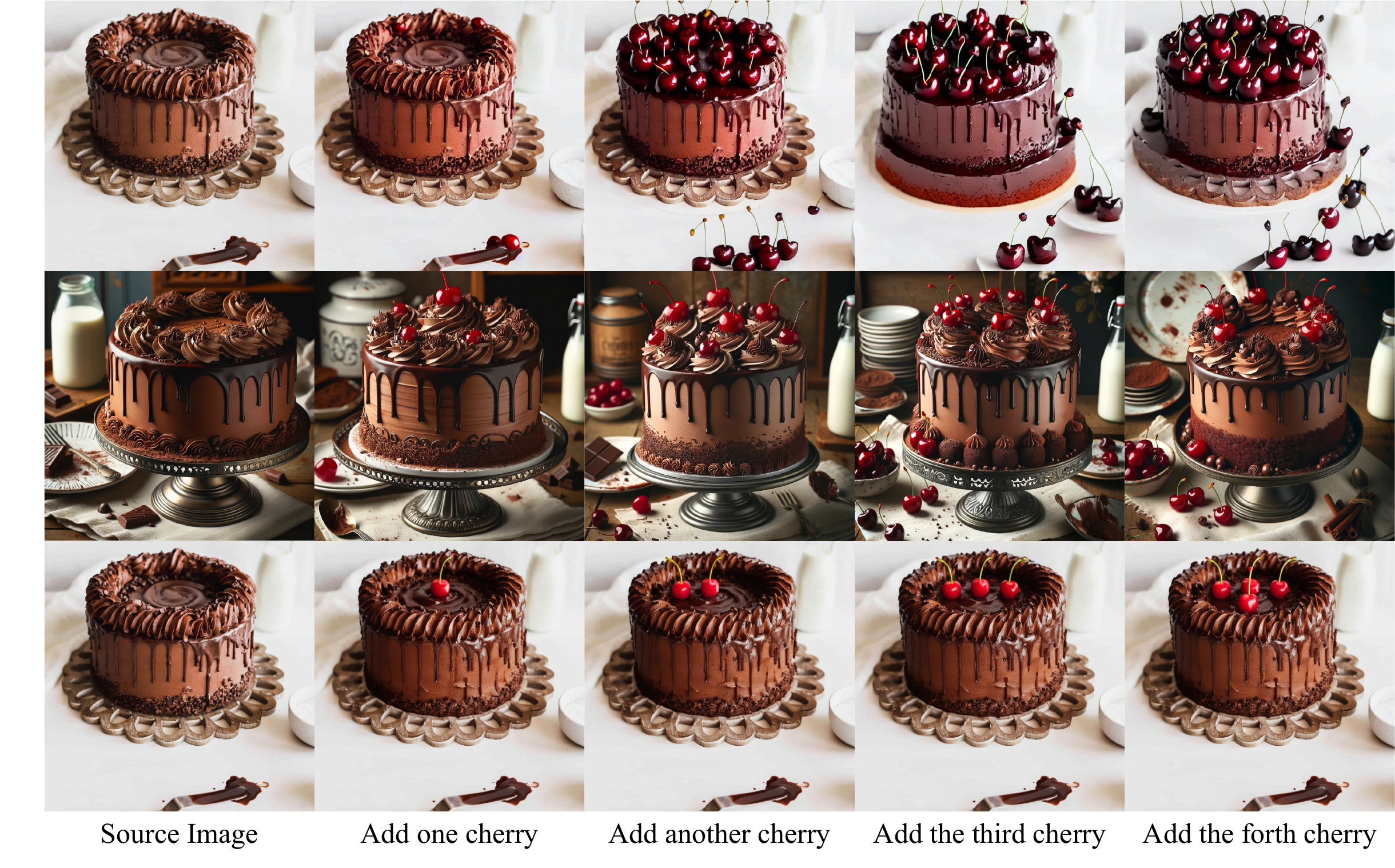}
    \put(0,27) {\rotatebox{90}{\tiny Ours}}
    \put(0,64) {\rotatebox{90}{\tiny DALL·E~3~\cite{dalle3}}}
    \put(0,105) {\rotatebox{90}{\tiny InstructPix2Pix~\cite{instructpix2pix}}}
    \end{overpic}
    \vspace{-1.65em}
    \caption{Comparisons with InstructPix2Pix~\cite{instructpix2pix} and DALL·E~3~\cite{dalle3} on serial addition. These text-based editing methods fail to follow precise and quantifiable instructions.}
    \label{fig:baseline_comparison_text}
    \vspace{-1em}
\end{figure}

\begin{figure*}[!t]
    \centering
    \includegraphics[width=\linewidth]{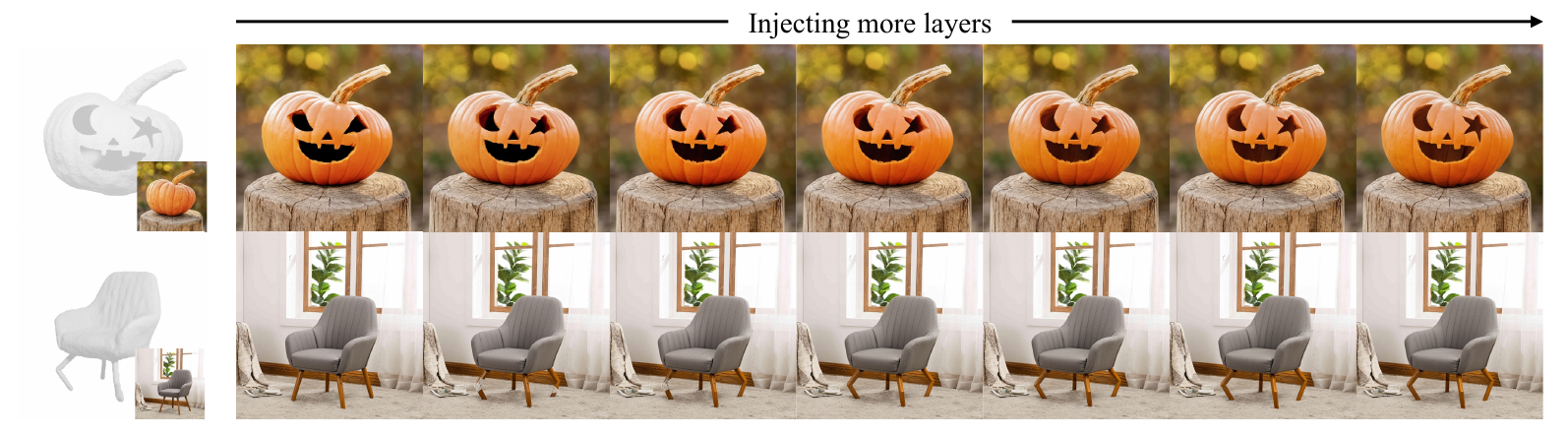}
    \vspace{-1.65em}
    \caption{Ablation studies on feature injection layers. From left to right, 
    progressively injecting more self-attention layers can result in increasingly improved alignment with user edits.}
    \label{fig:inject}
    \vspace{-1em}
\end{figure*}

\begin{table}[!t]
\small
\centering
\setlength\tabcolsep{4pt}
    \begin{tabular}{lccc}
        \toprule
        \textbf{Methods} & \textbf{DINO$\uparrow$} & \textbf{CLIP-I$\uparrow$} & \textbf{D-RMSE$\downarrow$} \\
        \midrule
        \textcolor{gray}{Original Coarse Rendering} & \textcolor{gray}{0.758} & \textcolor{gray}{0.902} & \textcolor{gray}{0.00} \\
        SDEdit~\cite{meng2021sdedit} ($t_0=0.4$) & 0.788 & 0.919 & 1.71 \\
        SDEdit~\cite{meng2021sdedit} ($t_0=0.6$) & 0.800 & 0.920 & 2.12 \\
        Ours w/o Feature Injection & 0.848 & 0.925 & 2.33 \\
        Ours w/o Depth Control & 0.851 & 0.921 & 2.15 \\
        \midrule
        Ours & 0.853 & 0.921 & 1.99 \\
        \bottomrule
    \end{tabular}
    \caption{Ablation studies of the enhancement methods on \emph{SculptingBench}. DINO score and CLIP-I score measure the textural details, and D-RMSE measures the geometric fidelity. We observe that depth control and feature injection can significantly enhance texture quality while maintaining geometric consistency.}
    \label{tab:ablation}
    \vspace{-1em}
\end{table}

\thinparagraph{Qualitative Results}
We showcase qualitative results in Fig~\ref{fig:results}, covering six precise image editing tasks. Detailed descriptions of these tasks are presented in Appendix~\ref{app:implementation_details}. Qualitatively, our method combines the creative freedom of generative models with the precision of graphics pipelines to achieve precise, quantifiable, and physically plausible outcomes for object editing across a variety of scenarios.

Our approach introduces new editing features through precise 3D geometry control, a capability not present in existing methods. We compare our method with the state-of-the-art object editing techniques for a comprehensive analysis. In Fig~\ref{fig:baseline_comparison_obj3dit}, we show that 3DIT~\cite{3dit}, designed for 3D-aware editing via language instructions, faces limitations when applied to real, complex images, largely because its training is based on a synthetic dataset. In Fig~\ref{fig:baseline_comparison_pose}, we compare the pose editing ability with DragDiffusion~\cite{dragdiffusion} and ControlNet~\cite{controlnet}. This comparison reveals that these methods encounter difficulties with complex pose manipulations because they are constrained to the 2D domain. Furthermore, in Fig~\ref{fig:baseline_comparison_text}, we show how text-based editing methods like InstructPix2Pix~\cite{instructpix2pix} and DALL·E~3~\cite{dalle3} struggle with precise and quantifiable instructions.

\thinparagraph{Ablation Studies}
\label{sec:ablation}
We create a new dataset \emph{SculptingBench} to evaluate our new image editing capabilities. This dataset contains 28 images covering six categories: pose editing, rotation, translation, composition, carving, and serial addition (see Appendix~\ref{app:sculptingbench}). We perform quantitative studies using different coarse-to-fine enhancement methods. To measure the visual similarity between the edited images and the original ones, particularly in terms of maintaining textural details through the editing process, we employ DINO and CLIP-I scores~\cite{dreambooth} as our metrics. To evaluate the geometric fidelity of user edits after enhancement, we introduce a novel metric, \emph{D-RMSE}. 
This metric is specifically created to evaluate how well geometric information is retained after the enhancement procedure. \emph{D-RMSE} measures the discrepancies between the depth maps of the coarse renderings and their enhanced counterparts:
\begin{equation*} \small \setlength\abovedisplayskip{6pt}
\text{D-RMSE} = \sqrt{\mathbb{E}\left[(\text{depth}_{\text{coarse}} - \text{depth}_{\text{enhanced}})^2\right]}
\end{equation*}
where $\text{depth}_{\text{coarse}}$, $\text{depth}_{\text{enhanced}}$ denote the depth maps MiDaS ~\cite{ranftl2020towards} estimates, for the coarse rendering and the enhanced output image, respectively. In Table~\ref{tab:ablation}, we show that without any enhancement, the textural quality metrics (DINO and CLIP-I scores) are quite low. SDEdit effectively preserves the edited geometry with a low D-RMSE, yet the visual quality significantly deteriorates compared to the original image (see Fig.~\ref{fig:c2f_comparison}). Our method offers a more advantageous balance, significantly enhancing texture quality as demonstrated by higher DINO and CLIP-I scores, while preserving geometric consistency, evidenced by a low D-RMSE score. 
We observe that both feature injection and depth control contribute to enhanced geometric consistency and can lead to further improvement when used together. 
Additionally, we conduct an empirical study to explore the ideal number of self-attention layers for injection. Fig~\ref{fig:inject} shows that more layers improve alignment with user edits. In our work, we use all layers for injection.

\section{Limitation}
Our method is an initial step towards integrating traditional geometric processing with advanced diffusion-based generative models for precise object editing.
Yet, it has limitations. 
A significant challenge is the dependency on the quality of single-view 3D reconstruction, which is anticipated to improve over time. Additionally, mesh deformation often requires some manual efforts for model rigging. 
Future research might explore data-driven techniques~\cite{Li:2021} to automate this process. The output resolution of our pipeline also falls short of industrial rendering systems, and incorporating super-resolution methods could be a solution for future improvements. Another issue is the lack of background lighting adjustment, which undermines the realism of the scene; future work could benefit from integrating dynamic (re-)lighting techniques. We present some instances of failure our our system in Appendix~\ref{app:failure_cases}.

\paragraph{Acknowledgements.} We thank Ellis Brown, Fred Lu, Sanghyun Woo, Adithya Iyer and Oscar Michel for helpful discussions. The research is partly supported by Intel, Cirrascale and the Google TRC program.

{
    \small
    \bibliographystyle{ieeenat_fullname}
    \bibliography{main}
}

\clearpage
\appendix

\section{Generative Enhancement Details}
\label{alg:1}
\begin{algorithm}
\small
\caption{Generative Enhancement}
\SetAlgoLined
\textbf{Define:} Pre-trained 2D text-to-image diffusion model $M$, input image $I$, coarse image $I_c$, enhanced image $I_f$, inversion prompt $y_{inv}$, prompt $y$, depth map $D$
\vspace{0.1cm}\\
$\hat{M} \leftarrow $\textsc{Fine-tune DreamBooth}($I$) \\
$\boldsymbol{x}_0^c, \hdots \boldsymbol{x}_T^c \gets$ \textsc{DDIM-Inversion}($I_c$, $y_{inv}$, $\textcolor{black}{D}$; $\hat{M}$)\; 
$\boldsymbol{x}_T^f\gets \boldsymbol{x}_T^c$ \\
\For{$t \gets T$ \KwTo $0$}{
    $\boldsymbol{f}_t^c, \boldsymbol{A}_t^c \gets \epsilon_\theta(\boldsymbol{x}_t^c, {y_{inv}}, \textcolor{black}{D}, t; \hat{M})$\\
    $\boldsymbol{f}_t^f, \boldsymbol{A}_t^f \gets \epsilon_\theta(\boldsymbol{x}_t^f, y, \textcolor{black}{D}, t; \hat{M})$ \\
    \textbf{if} $t > \tau_f$ \textbf{then} $\boldsymbol{f}_
    t^f \gets \boldsymbol{f}_t^c$ \\
    \textbf{if} $t > \tau_A$ \textbf{then} $\boldsymbol{A}_t^f \gets \boldsymbol{A}_t^c$ \\
    $\boldsymbol{\epsilon}_{t-1}^f \gets \epsilon_\theta(\boldsymbol{x}_t^f, y, \textcolor{black}{D}, t; \boldsymbol{f}_t^f, \boldsymbol{A}_t^f; \hat{M})$ \\
    $\boldsymbol{x}_{t-1}^f \gets$ \textsc{DDIM-Denoising}($\boldsymbol{x}_t^f, \boldsymbol{\epsilon}_{t-1}^f$; $\hat{M}$)
}
\KwResult{$I_f \gets$ \textsc{decoder}($\boldsymbol{x}_0^f$)}
\end{algorithm}
The generative enhancement pipeline starts with fine-tuning DreamBooth~\cite{dreambooth} with the input image. Subsequently, we apply depth control DDIM inversion~\cite{song2020denoising} to the coarse rendering image. The prompt $y_{inv}$, which describes the coarse rendering, is used to obtain the inverted latent for each time step. During each denoising step, we denoise the inverted latent of the coarse rendering and the latent of the refined image, extracting their respective feature maps, $\boldsymbol{f}_t^c$ and $\boldsymbol{f}_t^f$, as well as their self-attention maps $\boldsymbol{A}_t^c$ and $\boldsymbol{A}_t^f$. This step is formulated as:
\begin{align*}
    (\boldsymbol{f}_t^c, \boldsymbol{A}_t^c) &\leftarrow \epsilon_\theta(\boldsymbol{x}_t^c, y_{inv}, D, t) \\
    (\boldsymbol{f}_t^f, \boldsymbol{A}_t^f) &\leftarrow \epsilon_\theta(\boldsymbol{x}_t^f, y, D, t)
\end{align*}
Here, $\epsilon_\theta(\cdot)$ is the text-to-image diffusion model, specifically in our context, the Stable Diffusion XL ~\cite{podell2023sdxl} model. For the coarse rendering, the latent is denoted by $\boldsymbol{x}_t^c$, the inversion prompt by $y_{inv}$, and the depth map by $D$. For the refined image, the latent is represented by $\boldsymbol{x}_t^f$, and the prompt by $y$. 
Following Plug-and-Play~\cite{pnp}, we replace the feature and self-attention maps of the enhanced image with those from the coarse input:
\begin{align*}
    \boldsymbol{\epsilon}_{t-1}^f = \epsilon_\theta(\boldsymbol{x}_t^f, y, D, t; \boldsymbol{f}_t^f, \boldsymbol{A}_t^f)
\end{align*}
Here $\epsilon_\theta(\cdot; \boldsymbol{f}_t^f, \boldsymbol{A}_t^f)$ represents the model with replaced feature and self-attention maps, and $\boldsymbol{\epsilon}_{t-1}^f$ is the prediction for the refined image. Replacement stops once the current time step is below the thresholds $\tau_f$ and $\tau_A$. The threshold is important because the feature/self-attention maps may contain undesired artifacts from coarse 3D reconstruction and mesh deformation. 
\section{Implementation Details}
\label{app:implementation_details}
In this section, we provide implementation details of all our 6 tasks.
\thinparagraph{Pose Editing}
Pose editing is carried out by manually creating a skeleton for each 3D model and computing its skinning weights~\cite{baran2007automatic}. The object's pose is edited by adjusting the skeleton's bones. A text prompt is not required to describe the pose. 

\thinparagraph{Rotation}
Rotation is achieved by spinning the 3D model around its centroid. This allows us to rotate the model at any angle and then render it back into a 2D image. However, it becomes challenging to discern the viewpoint (\eg, front, back, or side), given only the coarse rendering image. Optional text prompts are helpful in guiding the denoising step and preventing the Janus Problem. If the rotation angle ranges from $[-45^\circ, 45^\circ]$, we add ``\textit{front view}'' to the text prompt. For angle between $[135^\circ, 225^\circ]$, we append ``\textit{back view}'' to the text prompt. For all other angles, we use ``\textit{side view}''. 

\thinparagraph{Translation}
Translation can be done by moving the 3D model within the 3D space in any direction and over any specific distance. As the translated model is rendered, the camera perspective adjusts accordingly.  As illustrated in Fig~\ref{fig:results}, moving the dog or the truck closer to the camera results in an enlarged image of the object, consistent with the camera's perspective.

\thinparagraph{Composition}
Our method allows for the addition of artist-created 3D objects to the scene. In Fig~\ref{fig:results}, we insert various models into the scene. Despite the 3D models not being of high quality, our coarse-to-fine strategy significantly enhances their detail, as evident in the tiger example where the texture displays hair details and a realistic face in the final output, blending well with the environment. Note that these models are not used for fine-tuning during DreamBooth training. In certain cases, text prompts prove helpful in guiding the denoising step and supplementing our geometric guidance. 

\thinparagraph{Carving}
Beyond mesh deformation, our method enables cutting and removing parts of the mesh through the use of molds. In Fig~\ref{fig:results}, a moon-and-star-shaped mold is positioned against a pumpkin's surface. By calculating and excising the overlapping areas, the resulting mesh resembles a finely carved pumpkin in specific shapes.

\thinparagraph{Serial Addition}
Similar to composing elements, we can take meshes reconstructed from images and integrate them into the scene one by one. In Fig~\ref{fig:results}, we adjust each fish and duck's size, pose and their orientation before adding it to the scene. Our approach realistically merges the coarse 3D fish models into the scene, maintaining a realistic appearance even with reflections on the water's surface.

\begin{figure*}[!t]
    \centering
    \includegraphics[width=0.81\linewidth]{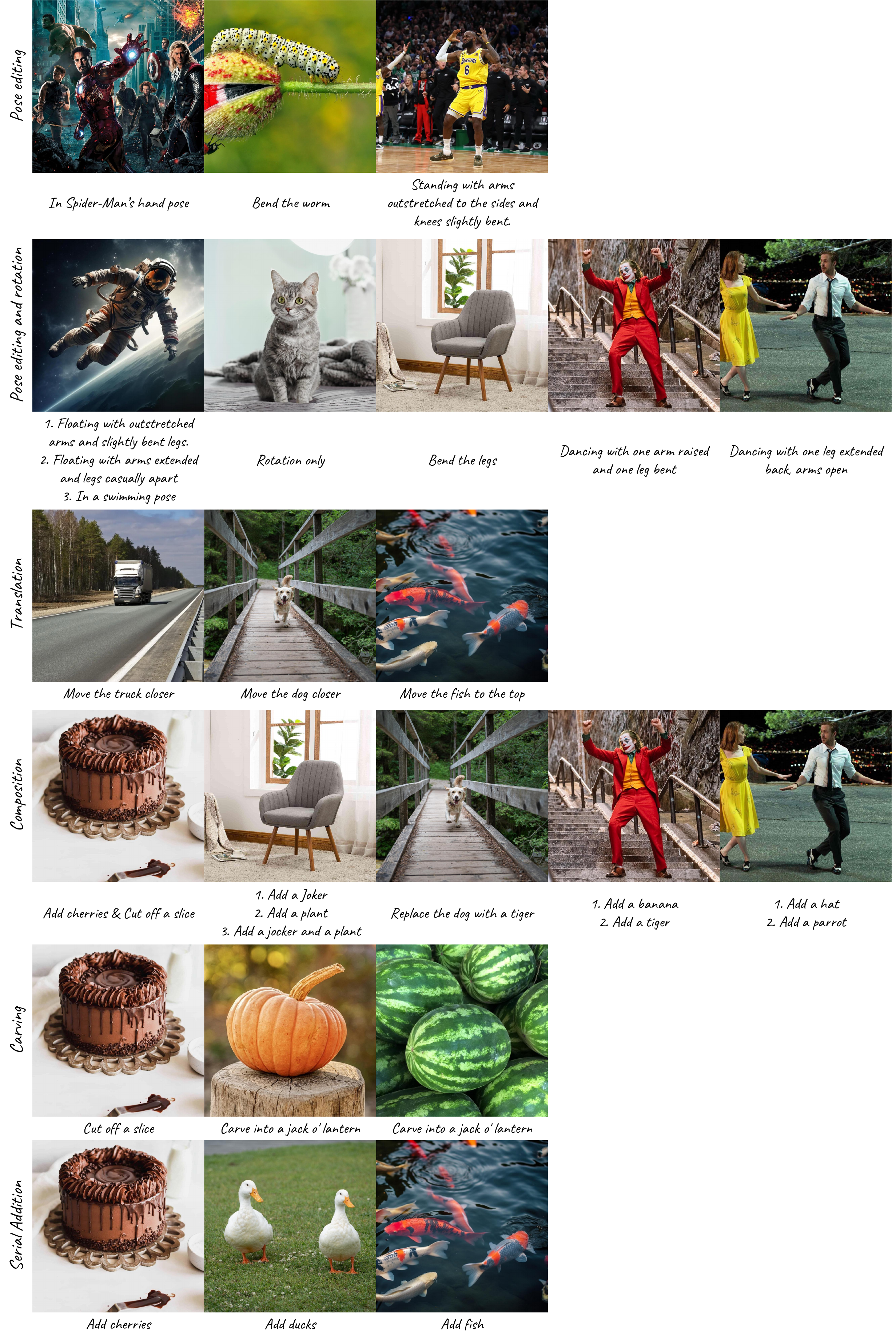}
    \caption{All 28 edits and 15 input images of our \emph{SculptingBench}. We provide textual descriptions of the edits here. However, in practice we aim to make precise, quantifiable edits directly to 3D models, without relying on text prompts.}
    \label{fig:sculptingbench}
\end{figure*}

\section{\emph{SculptingBench}}
\label{app:sculptingbench}
Our \emph{SculptingBench} dataset comprises 28 edits applied to 15 images, encompassing each of the 6 editing tasks we have developed. The full dataset is illustrated in Fig.~\ref{fig:sculptingbench}. These instances present significant challenges to current object editing techniques, thereby serving as an ideal platform for testing and developing precise object editing methods.

\section{Failure Cases}
\label{app:failure_cases}
One key limitation lies in its dependence on the quality and reliability of single-view reconstruction techniques, particularly when dealing with unseen perspectives. Any errors in this process can result in editing failures. 

\begin{figure}[!h]
    \centering
    \includegraphics[width=\linewidth]{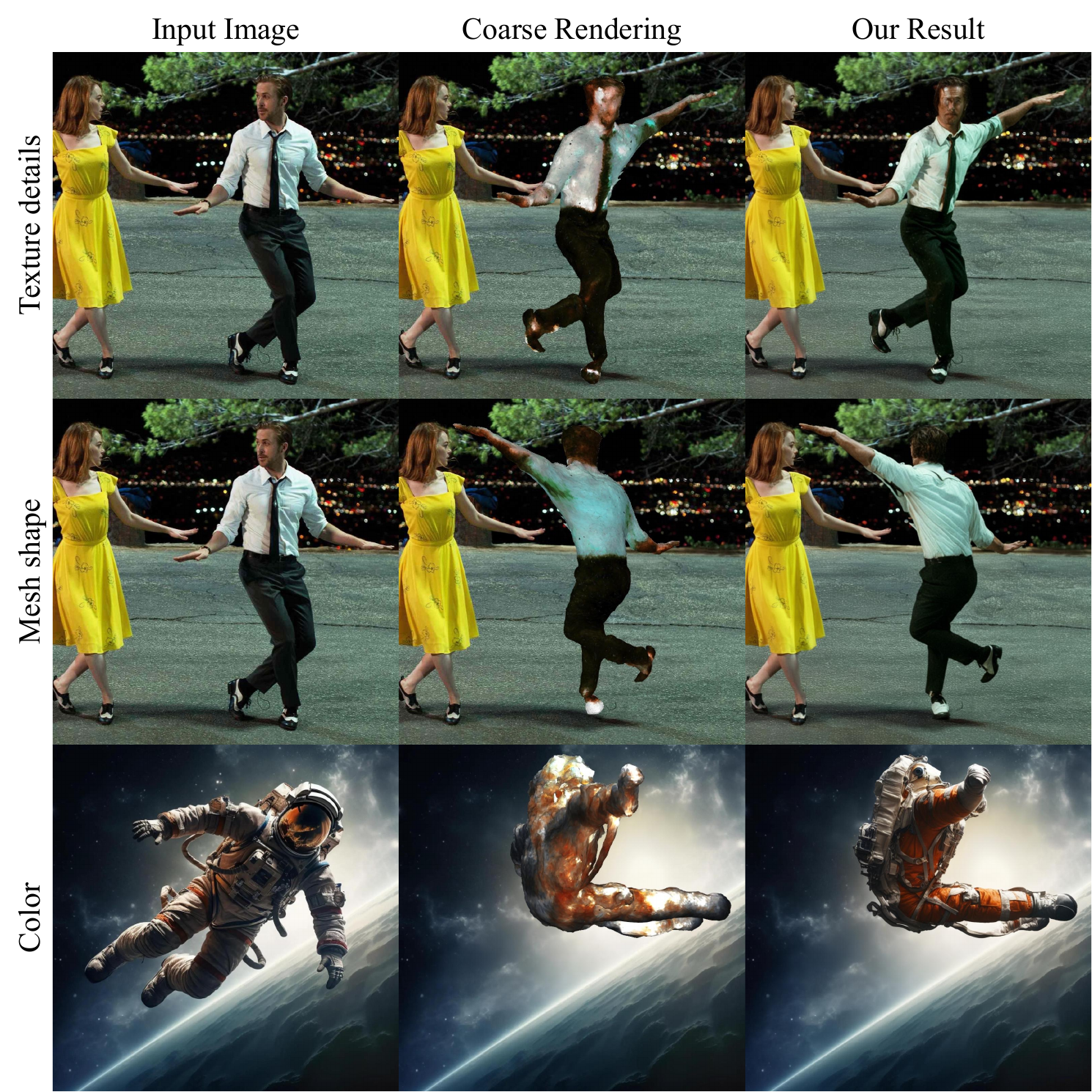}
    \caption{Failure cases due to inaccurate reconstruction of texture, geometry, and color.}
    \label{fig:failure}
\end{figure}

As demonstrated in Fig~\ref{fig:failure}, the reconstruction occasionally fails to produce detailed textures, leading to a blurred face in the top row example. Challenges also arise in mesh reconstruction and extraction. The middle row displays artifacts beneath the man's armpit, stemming from imprecise reconstruction in that region. In the bottom row example, wrong color reconstruction resulted in an less realistic final color in the output.

\end{document}